\documentclass[reprint,amsmath,amssymb,aps,prb,floatfix, bibliography]{revtex4-1}
\usepackage{natbib}
\usepackage{units}
\usepackage{tabto}
\usepackage{color}
\usepackage{textcomp}
\usepackage{natmove}
\usepackage{mciteplus}
\usepackage[export]{adjustbox}
\usepackage{amsbsy}
\usepackage{amssymb}
\usepackage{subfloat}
\usepackage{float}
\usepackage{geometry}
\usepackage{setspace}
\usepackage{xkeyval}
\usepackage{graphicx}
\usepackage[caption=false]{subfig}
\usepackage{placeins}
\usepackage{tikz}
\usepackage{csquotes}
\usepackage{colortbl}
\usepackage{amsmath}
\usepackage{placeins}
\usepackage{graphicx}
\usepackage[unicode=true, bookmarks=true, bookmarksnumbered=false, bookmarksopen=false,
  breaklinks=false,pdfborder={0 0 0},backref=false,colorlinks=true]{hyperref}
 \hypersetup{pdftitle={Topological phases in hydrogenated gallenene and in its group elements},
  pdfauthor={Ritesh Kumar, Ranjan Kumar Barik and Abhishek K. Singh},
  pdfsubject={Computational Material Science},colorlinks=true,
  unicode=false,pdftoolbar=true,pdfmenubar=true,pdffitwindow=true, pdfstartview={FitH}, pdfnewwindow=true, pdfcreator={Latex}, linkcolor=red, citecolor=blue, urlcolor=blue}

\DeclareGraphicsExtensions{.jpg, .eps, .png}

\definecolor{NLRed}{RGB}{215,24,30}
\definecolor{abstractcolor}{RGB}{255,243,201}

 \makeatother

\begin{document}

\title{Topological phases in hydrogenated gallenene and in its group elements}

\author{Ranjan Kumar Barik}
\author{Ritesh Kumar}
\author{Abhishek K. Singh}
\email[Corresponding Author: ]{abhishek@iisc.ac.in}
\affiliation{Materials Research Centre, Indian Institute of Science, Bangalore 560012, India}

% \abbreviations{DFT, TSM, TP}
% \keywords{topological materials, triple-point fermion, nexus fermion, nodal line}

\begin{abstract}
Nontrivial topology of Dirac (DSMs), Weyl (WSMs) and nodal line semimetals (NLSMs) are delineated by the novel band crossings near the Fermi level in the bulk and the appearance of exotic surface states. Among them, nodal line semimetals have gained immense interest due to the formation of one-dimensional nodal ring near the Fermi level. Using density funtional theory (DFT) calculations, we report that two dimensional (2D) NLSM phase can be hosted on hydrogen passivated (010) surface of gallium (gallenene) and on other group 13 elements, without inclusion of spin-orbit coupling (SOC). NLSM in these 2D systems is protected by the presence of crystalline (CS) along with inversion (IS) and the time reversal symmetry (TRS). In the presence of SOC, aluminane preserved its topological NLSM phase while in other single-layered group 13 elements, a gap opened at the nodal point due to relatively stronger SOC effect. On applying tensile strain along with the inclusion of SOC, hydrogen passivated gallenene (gallenane) evolves into quantum spin Hall insulator with an indirect bandgap of 28 meV. The appearance of long range dissipationless linearly dispersive helical edge states and a large gap in gallenane make it promising for room temperature spintronics applications.
\end{abstract}

\maketitle

\section{Introduction}
The herald of two dimensional (2D) and three dimensional (3D) topological insulators (TI)\cite{PhysRevLett.95.226801,Bernevig1757,PhysRevLett.98.106803,RevModPhys.82.3045, moore2010birth} have paved new areas of research in topological science. These insulating phases of matter have a gapless edge (2D) or surface (3D) states, protected by the nontrivial topological behaviour of the bulk wavefunction. In recent times topological semimetals (TSMs) \cite{PhysRevB.83.205101, PhysRevLett.107.186806, PhysRevB.84.235126, 
PhysRevLett.107.127205, PhysRevLett.108.266802, PhysRevLett.108.046602, PhysRevLett.108.140405, PhysRevLett.113.027603, PhysRevLett.115.036807} namely, DSMs, WSMs and NLSMs have attracted more attention due to their exotic behaviour of bulk as well as surface states. DSMs and WSMs have zero dimensional node having four-fold and two-fold band degeneracies, respectively in the bulk together with topologically protected arc-like surface states\cite{PhysRevB.92.115428,PhysRevLett.116.066802, Kargarian8648}.  Beyond two-fold and four-fold degeneracies, fermionic systems having three, six and eight fold degeneracies have also been identified recently\cite{multiple-tpf, Bradlynaaf5037}. In NLSMs, one-dimensional band crossings give rise to a continuous closed loop, called nodal ring, which is protected by the mirror symmetry in addition to TRS and IS\cite{PhysRevB.97.241111, PhysRevB.93.205132}. Investigating the non-trivial topology of these gapless materials in 2D systems has become one of the most exciting frontiers of research, due to possible applications in nano-electronic devices such as photodetector\cite{yan2014topological}, field-effect transistor\cite{vandenberghe2014realizing}, topological quantum computers\cite{kou2017two}. Two dimentional NLSMs have been predicted in several systems such as in $B_{2}$C\cite{1710.05144}, $C_{9}N_{4}$\cite{C8TA02555J}, square and Honeycomb-Kagome lattice\cite{0256-307X-34-5-057302, PhysRevLett.111.130403}, $X_{2}Y$ (X = Ca, Sr, and Ba; Y = As, Sb, and Bi)\cite{PhysRevB.95.235138}. In most of the cases, NLSMs are observed only when SOC is not considered whereas with the inclusion of SOC, NLSMs either become strong TI\cite{PhysRevB.95.045136} or evolve to a stable Dirac point\cite{PhysRevLett.115.036807}. 

Recently, it was shown theoretically that 2D borophene (a group 13 element) exhibited exotic topological behaviour (DSM) when passivated completely by hydrogen\cite{PhysRevB.96.195442}. Gallenene, the 2D counterpart of bulk gallium (also a group 13 element), was synthesized very recently and its electronic properties were studied from first-principles calculations\cite{Kochate1701373}. The study revealed highly dispersive bands with partially-filled Dirac cones. The potential applications of gallenene in electrical connector and thermal barrier in devices was also proposed. However, its topological properties have not been studied yet.
Therefore, we have investigated topological features of 2D hydrogenated forms of heavier group 13 elements such as aluminium, gallium, indium and thallium. All of them exhibit NLSM phase when SOC is neglected, which is protected by mirror symmetry along with TRS and IS. We find that NLSMs exist in aluminane even in the presence of SOC, whereas in all other hydrogenated 2D structures, a small gap is observed at the nodal points due to relatively stronger SOC strength. Interestingly, we have found that gallenane evolves into quantum spin Hall (QSH) insulator or 2D TI with a indirect gap of 28 meV under uniaxial tensile strain of $-$2.6$\%$ along a-direction. The large dispersive helical egde states of gallenane separates it from other QSH insulator which can be used for quantum transport.

\section{\label{sec:theory}Methodology}

First-principles calculations were performed using the density functional theory (DFT), implemented in the Vienna ab initio simulation package (VASP)\cite{PhysRevB.47.558} code. The projector augmented wave (PAW) \cite{PhysRevB.59.1758} potentials were used to describe the ion-electron interactions. The exchange-correlation effects were approximated by Perdew-Burke-Ernzerhof (PBE) functional under the generalized gradient approximation (GGA)\cite{PhysRevLett.77.3865}. An energy cut-off of 500 eV was used to expand the wavefunctions in a plane wave basis. A $\Gamma$-centered 15$\times$15$\times$1 Monkhorst-Pack \cite{PhysRevB.13.5188} K-point grid was used to sample the Brillouin zone. The structures were optimized by employing a conjugate gradient scheme until each component of Hellmann-Feynman forces on the atoms were less than 0.005 eV/\AA. Spin-orbit coupling was also included for the relativistic effects. The Bloch spectral functions and Fermi surfaces were calculated using iterative Green's function method\cite{Sancho1985} by using tight-binding Hamiltonian from the maximally localized Wannier functions (MLWFs)\cite{mostofi2014updated}, implemented in the WannierTools package\cite{WU2018405}. These MLWFs were generated from wannier90 code\cite{wannier90}.

\section{\label{results-and-discussion}Results and Discussion}

\begin{figure}
\begin{center}
\includegraphics[width=\columnwidth]{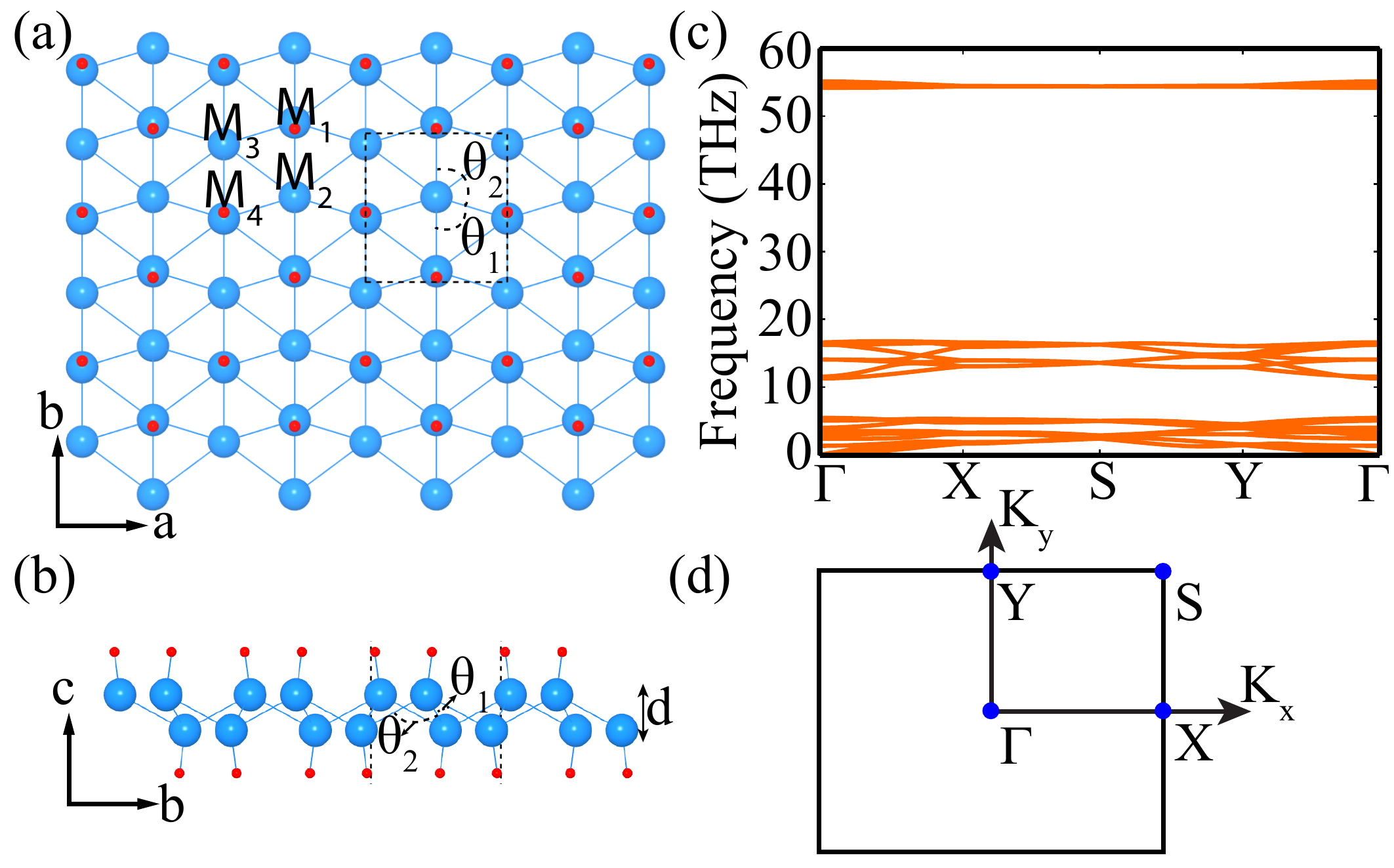}
\end{center}
\caption{ shows (a) top and (b) side view of gallenane (c) phonon bandstructure of gallenane and (d) rectangular Brillouin zone with high-symmetric path.} 
\label{fig:crystall structure}
\end{figure}

%\subsection{Crystal structure and stability}
$\alpha$-Gallium is the most stable phase among the all existing bulk phases. It exists in the space group \textit{Cmce} (${D_{2h}^{18}}$, No-64). Its exfoliation was carried out along (010) and (100) directions, however, the metal-covalent mixed interactions in (010) layer implied that this layer is more stable than the (100) gallium layer, which is also observed experimentally\cite{Kochate1701373}. Without hydrogen passivation, gallenene shows metallic nature and has no interesting topological properties. However hydrogen passivated gallenene, we called it as gallenane, exhibits NLSM in the absence of SOC. 
%However, NLSM is exhibited by passivating gallenene with hydrogen, we called it as gallenane. 
Fig.\protect\ref{fig:crystall structure} (a) shows the top view of gallenane and the black dash line represents the primitive cell, containing four Ga-atom and four H-atom. It has two layers as shown in Fig.\protect\ref{fig:crystall structure} (b) and belongs to space group \textit{Pbcm} or \textit{P2/b $2_{1}$/c $2_{1}$/m} (${D_{2h}^{11}}$ No-57). These layers can be seen as a puckered trigonal gallium network with each gallium atom bonded with one hydrogen atom. This structure is characterized by two angles, ${\theta_{1}}$ and ${\theta_{2}}$ and three M$-$M bond lengths along with buckling height as shown in Fig.\protect\ref{fig:crystall structure} (a) and (b). For gallenane, ${\theta_{1}}$ is 62.97$^\circ$,  ${\theta_{2}}$ is 75.37$^\circ$ and three M$-$M bond lengths are ${d_{M_{1}-M_{2}}}$ = 2.71 \AA, ${d_{M_{2}-M_{3}}}$ = 2.83 \AA, ${d_{M_{2}-M_{4}}}$ = 2.71 \AA, with buckling height d = 1.3 \AA. The lattice parameters of gallenane are, a = 4.55 {\AA} and b = 4.77 \AA. Structural information of other 2D analogous group 13 elements  are provided in supporting information. The dynamical stability of all systems are verified by phonon calculations. Fig.\protect\ref{fig:crystall structure} (c) shows the phonon dispersion of gallenane which does not contain any imaginary frequencies, signifies that our system is dynamically stable. Fig.\protect\ref{fig:crystall structure} (d) display the rectangular brillouin-zone with high symmetric points.

\begin{figure}
\begin{center}
\includegraphics[width=\columnwidth]{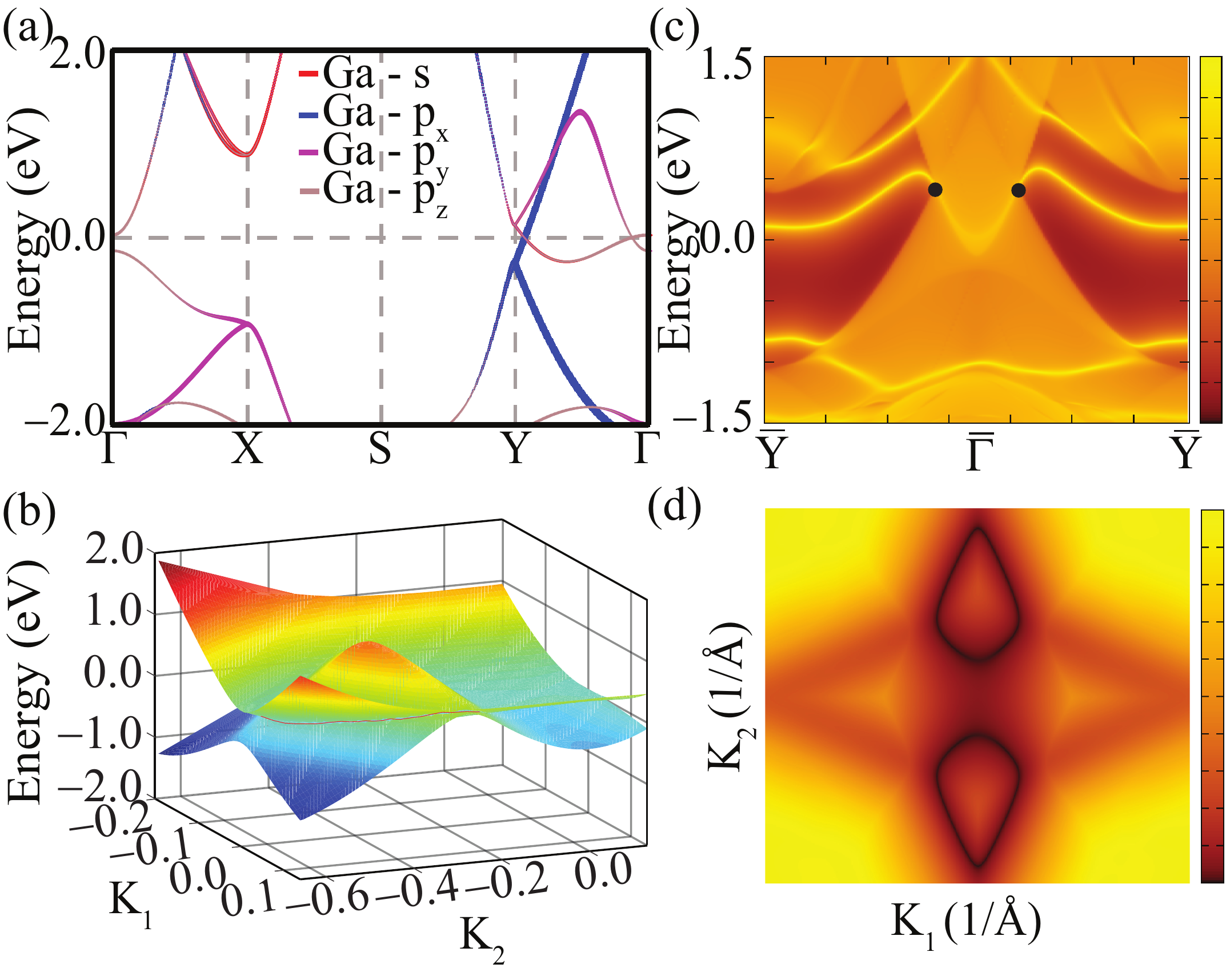}
\end{center}
\caption{describes (a) orbital resolved band-structure of gallenane without inclusion of SOC (b) nodal-line semimetal in 3D view (c) drum head state with black circle representing two nodal points in the bulk as $NP_{1}$, $NP_{2}$ and (d) projection of gap spectral function of nodal ring on k-plane} 
\label{fig:without soc}
\end{figure}

%\subsection{Gallenane without SOC}
Electronic band structure of gallenane computed in the absence of SOC as shown in the Fig.\protect\ref{fig:without soc}(a) with Fermi-level is set to zero. Along Y-$\Gamma$, two bands are crossing linearly at 54 meV above and 21 meV below the Fermi-energy, forming a doubly-degenerate (Weyl) NLSM. These bands are mostly contributed from Gallium s and p orbitals. The nontrivial topology in gallenane can be examined by bulk-boundary correspondence (BBC) principles, which says that (d-1)-dimensional edge state is protected by topology of d-dimensional bulk. The drumhead in the BBC state is the main signature of nontrivial topology in NLSM\cite{PhysRevB.93.121113}. Direct evaluation of BBC states using DFT calculation is computationally very expensive. Therefore, to examine the topological nodal states, we have calculated the band structure for gallenane, based on the tight-binding model constructed by maximally localized Wannier functions method. Fig.\protect\ref{fig:without soc} (b) show the NLSM in three dimensional view where the gap in the bulk at the Fermi level closes along a loop. Using the iterative Green's function method, we have computed a momentum-dependent surface Green function (SGF) for a semi-infinite system with edge along x-direction\cite{WU2018405}. The edge spectral function can be calculated from imaginary part of SGF as: 

%\begin{equation}
\[A(k_{\parallel},\omega)= -\frac{1}{\pi} \lim_{\eta\to 0^+} \emph{Im}Tr{G_{s}}(\mathbf{k}_{\parallel}, \omega + i\eta)\]
%\label{eqn:bloch-spectral-function}
% \end{equation}
where, the SGF is given by ${G_{s}}(\mathbf{k}_{\parallel}, \omega + i\eta) \simeq (\omega - \varepsilon_{n}^s)^{-1} $. 
Drum head state shown in Fig.\protect\ref{fig:without soc} (c), which comprises of bright yellow line connecting two nodal point (marked as black cirle). Fig.\protect\ref{fig:without soc} (d) showing the gap spectral function contains two closed loop due to the two-fold rotational symmetry.

%\subsection{Symmetry analysis}
Symmetry plays an important role in preserving topological phases, like in DSMs, Dirac node is protected by TRS and IS whereas in WSMs, Weyl node is protected if TRS or IS is broken. Moreover some extra degeneracies in electronic band structures could arise due to nonsymmorphic symmetries\cite{PhysRevLett.115.126803} . Nonsymmorphic symmetries $\{$g$|$t$\}$ are characterized by a point group operation g with a fractional translation t of Bravais lattice vector. 
%A glide operation defined by a mirror g = $M_{i}$ (where i = x, y, z) perpendicular to the i-axis followed by half lattice vector translation along the other directions. However, a screw operation defined by a rotation g = $C_{n}$  around the i-axis followed by fractional translation along the i-axis. 
Our system is a centro-symmetric non-magnetic system therefore it preserves both TRS and IS. In addition to that, this system have a b-glide represented by $\{${$M_{x}$}$|$b/2$\}$(x, y, z) = (-x, y+1/2, z), a c-glide represented by $\{${$M_{y}$}$|$c/2$\}$ = (x, y, z) = (x, -y, z+1/2) and a mirror $M_{z}$(x, y, z) = (x, y, -z). Also there exist three screw axes, which are mutually orthogonal to each other. The edges on $k_{x}$ = $\pi$ and $k_{y}$ = $\pi$ planes where two glide operations with glide plane orthogonal to each other anticommute. Therefore, all bands along the edges of brillouin zone are doubly degenerate as shown in X-S and S-Y path in Fig.\protect\ref{fig:without soc} (a). Now along Y-$\Gamma$, $k_{z}$ = 0 and $k_{x}$ = 0 planes intersect each other to form a nodal line. Since $k_{z}$ = 0  is a mirror plane and the two bands near the Fermi-level have opposite mirror eigenvalue, the band inversion gives rise to NLSM along Y-$\Gamma$ and that can be characterized by $\pi$ Berry phase. 

\begin{figure}
\begin{center}
\includegraphics[width=\columnwidth]{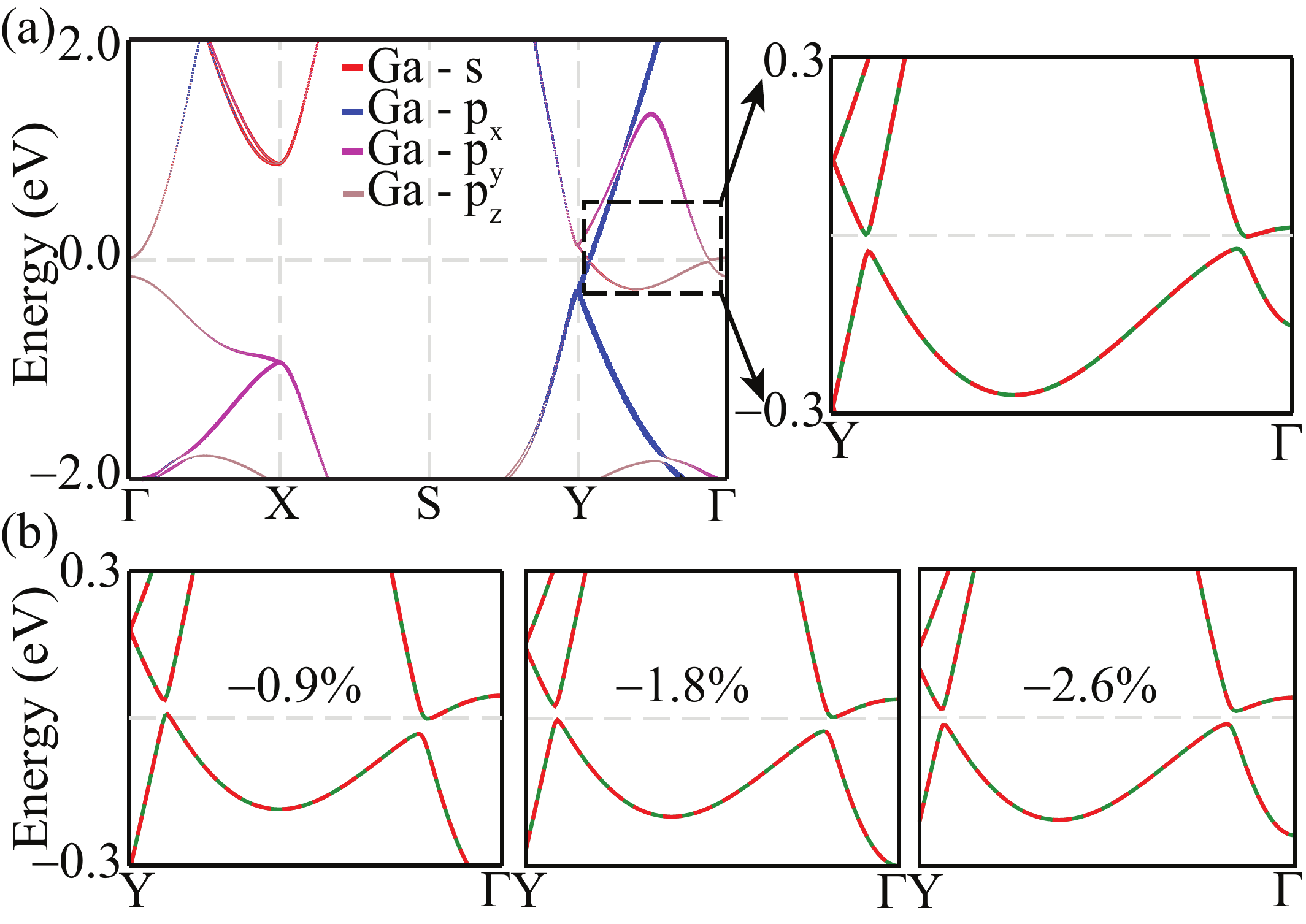}
\end{center}
\caption{shows (a) orbital resolved band structure of gallenane in the presence of SOC and (b) evolution of nodal gap under tensile strain along a-direction} 
\label{fig:strain}
\end{figure}

%\subsection{Gallenane with SOC}
 With the inclusion of SOC, bands become doubly degenerate along all high symmetry path, as shown in Fig.\protect\ref{fig:strain} (a). A tiny gap of 22 meV is induced in gallenane, but Fermi level crosses the conduction band by 2 meV as shown in zoomed Fig.\protect\ref{fig:strain} (a). To study the quantum spin Hall (QSH) insulating phase, it is required that Fermi level should lie within the gap. For that we applied both tensile and compressive uniaxial and biaxial external strain on gallenane. Under the tensile strain of $-$2.6$\%$ along a-direction, gallenane opened an indirect gap of 28 meV, with Fermi level lying in between the gap. Fig.\protect\ref{fig:strain} (b) shows the variation of gap at the nodal point with respect to external strain. 

 \begin{figure}
\begin{center}
\includegraphics[width=\columnwidth]{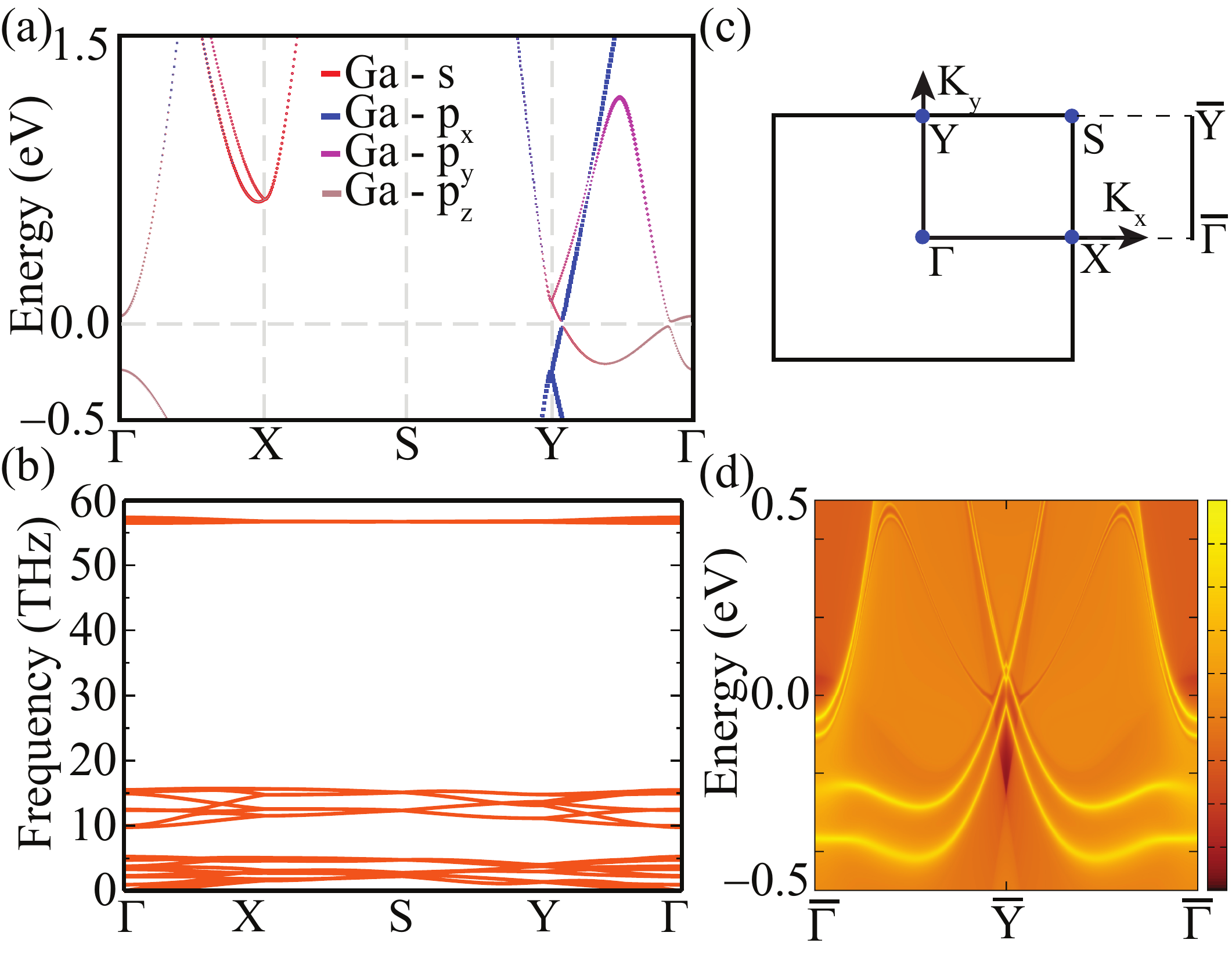}
\end{center}
\caption{showing (a) electronic band structure of gallenane under tensile strain of $-$2.6$\%$ along a-direction and its (b) phonon dispersion (c) (100) projection to evaluate edge states and (d) edge states of QSH insulator gallenane} 
\label{fig:qshi under strain}
\end{figure}

The full orbital-resolved band structure of gallenane at $-$2.6$\%$ strain along a-direction is shown in Fig.\protect\ref{fig:qshi under strain} (a). Along Y-$\Gamma$, a band inversion occurs between the $p_{x}$ and $p_{z}$ orbitals of gallium. This strained gallenane is also found to be dynamically stable, as evident from the phonon bandstructure shown in Fig.\protect\ref{fig:qshi under strain} (b). Nontrivial topology of gallenane at $-$2.6$\%$ tensile strain is confirmed by computing $Z_{2}$ topological invariant under the evolution of wannier center of charges (WCCs). We have taken (100) direction to compute edge state as shown in Fig.\protect\ref{fig:qshi under strain} (c). Helical edge state in BBC as shown in Fig.\protect\ref{fig:qshi under strain}(d) assured that gallenane is a QSH insulator under the tensile strain. These helical edge states are robust against non-magnetic disorder. The range of linear dispersion in the helical egde state of gallenane is of the order of 1 eV. Such a large dispersion in this potentially exfoliable gallenane monolayer may be efficient candidate for quantum transport.

The electronic bandstructures for other group 13 analogues of gallenane are shown in the supplementary information (Fig. S2 - S4). All of them exhibit a nodal ring consisting of two nodal points (NP$_1$ and NP$_2$) near the Fermi level like gallenane. While the position of NP$_1$ moves up with respect to the Fermi level, that of NP$_2$ shifts down, as we go from aluminane to thalinane. The gaps at NP$_1$ and NP$_2$ for all of these structures are tabulated in Table S2, calculated with and without the inclusion of SOC. From Table S2, we can infer that the gaps at NP$_1$ and NP$_2$ decrease on going from aluminane to thalinane when SOC is not included. However, the trend is reversed when the effect of SOC is included in the bandstructure calculations. While the two gaps at NP$_1$ and NP$_2$ close under SOC. It could be ascribed to the prominent relativistic effects in the heavier elements ,i.e., gallium, indium and thallium. Therefore, aluminane, once experimetntally realized, can be utilized in new electronic devices like topological current rectifier\cite{rui2018topological}.

\section{Conclusion}
In conclusion, we have theoretically predicted NLSMs in the 2D structures of heavier group 13 elements i.e. aluminane, gallenane, indinane and thalinane in the absence of SOC. Symmetry analysis implies that NLSM state is preserved due to the mirror symmetry along with TRS and IS. Presence of drumhead in the BBC states guarantee the bulk nontrivial topology in gallenane. With the inclusion of SOC, a very small gap is opened in gallenane, indinane and thalinane due to the strong SOC effects. However, aluminane still preserved its topological NLSM behaviour in the presence of SOC. With the tensile strain of $-$2.6$\%$ along the a-direction, a QSH insulating state is observed in gallenane. Large QSH insulating gap in gallenane assured that, it could be used in room temperature applications as low-power electronic and spintronic devices.

%\FloatBarrier

% \begin{suppinfo}
%\section{Suplemental Material}
%The phonon dispersion and band structures of materials are given in the supplemental material.
% \end{suppinfo}

\acknowledgements
We acknowledge Materials Research Center (MRC), Solid State Structural and Chemistry Unit (SSCU) and Supercomputer Education and Research Center (SERC), Indian Institute of Science for providing the computational facilities. 

\bibliography{manuscript}

\end{document}

\maketitle

\section{Crystal structure of gallenane analogous group 13 elements}
\begin{figure}[!htbp]
\centering
\includegraphics[width=1.1\columnwidth]{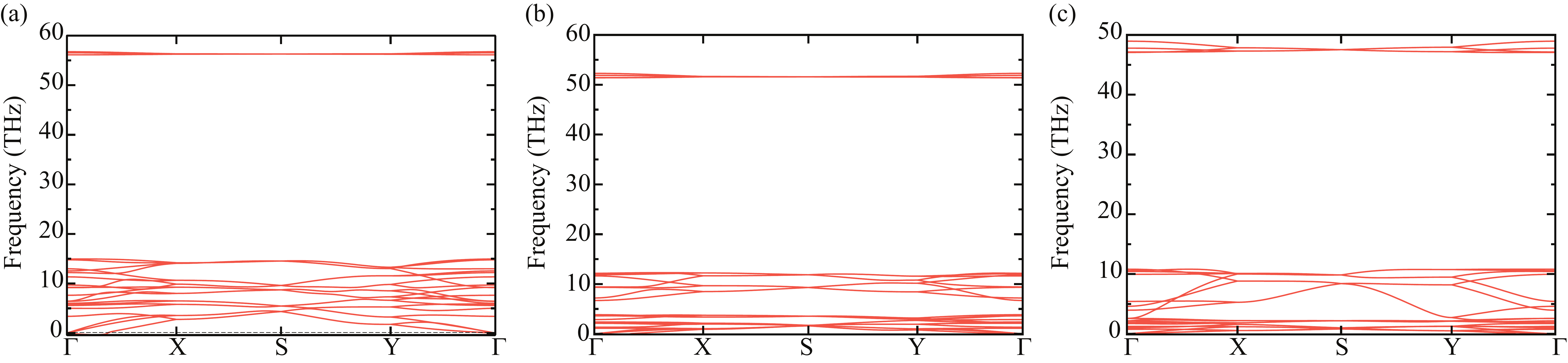} 
\caption{Phonon bandstructures of (a) aluminane (b) indinane (c) thalinane}
\label{figS1}
\end{figure}

Aluminane, indinane and thalinane are the analogous group 13  elements of gallenane. All these 2D structures have orthorhombic space group \textit{Pbcm}. The structural informations of all these systems are given in Table-\protect\ref{structure}. The M$-$M bond lengths in aluminane and indinane are same, however thalinane has larger M$-$M bond lengths as shown in Table-\protect\ref{structure}, coulmn 6$-$9. The interactions between two metal sites are lesser in thalinane as compared to that in aluminane and indinane. Also, hydrogen interacts more with aluminane and indinane as compared to thalinane as shown in Table-\protect\ref{structure}, coulmn 10. Dynamical stability of all structures are verified by the phonon bandstructures as shown in Fig-\protect\ref{figS1}.

\begin{table}[!htbp]
\caption{Structural parameters of hydrogenated group 13 monolayers} \label{structure}
\begin{center}
\begin{tabular}{| p{1.0cm} | p{1.0cm} | p{1.0cm} | p{1.5cm} | p{1.5cm} | p{1.5cm} | p{1.5cm} | p{1.5cm} | p{1.5cm} |}
\hline
\small{M} & \small{a (\AA)} & \small{b (\AA)} & \small{$\theta_1$ ($\degree$)} & \small{$\theta_2$ ($\degree$)} & \small{d$_{M_{1}-M{2}}$ (\AA)} & \small{d$_{M_{2}-M{3}}$ (\AA)} & \small{d$_{M_{2}-M{4}}$ (\AA)} & \small{d$_{Ga-H}$ (\AA)} \\ [0.3cm]
\hline
Al & 4.50 & 4.72 & 61.74 & 86.77 & 2.75 & 2.82 & 2.74 & 1.59 \\
In & 4.50 & 4.72 & 61.74 & 86.77 & 2.75 & 2.82 & 2.74 & 1.59 \\ 
Tl & 5.44 & 5.70 & 74.69 & 77.15 & 3.26 & 3.89 & 3.14 & 1.78 \\  
\hline
\end{tabular}
\end{center}
\end{table}

%\section*{Stability of hydrogenated group 13 monolayers}
\begin{table}[!htbp]
\caption{Gaps at nodal points in the bandstructures of
 hydrogenated group 13 monolayers} \label{nodal}
\begin{center}
\begin{tabular}{| p{1.5cm} | p{2.5cm} | p{3.0cm} | p{2.5cm} | p{3.0cm} |}
\hline
\small{M} & \small{$\Delta$E$_{NP_1}$ (PBE) [meV]} & \small{$\Delta$E$_{NP_1}$ (PBE-SOC) [meV]} & \small{$\Delta$E$_{NP_2}$ (PBE) [meV]} & \small{$\Delta$E$_{NP_2}$ (PBE-SOC) [meV]} \\ [0.3cm]
\hline
Al & 25.0 & 10.1 & 7.8 & 6.7 \\
In & 5.4 & 75.5 & 2.3 & 80.4 \\ 
Tl & 0.9 & 77.5 & 1.8 & 141.8  \\  
\hline
\end{tabular}
\end{center}
\end{table}

\begin{figure}[!htbp]
\centering
\includegraphics[width=1.0\columnwidth]{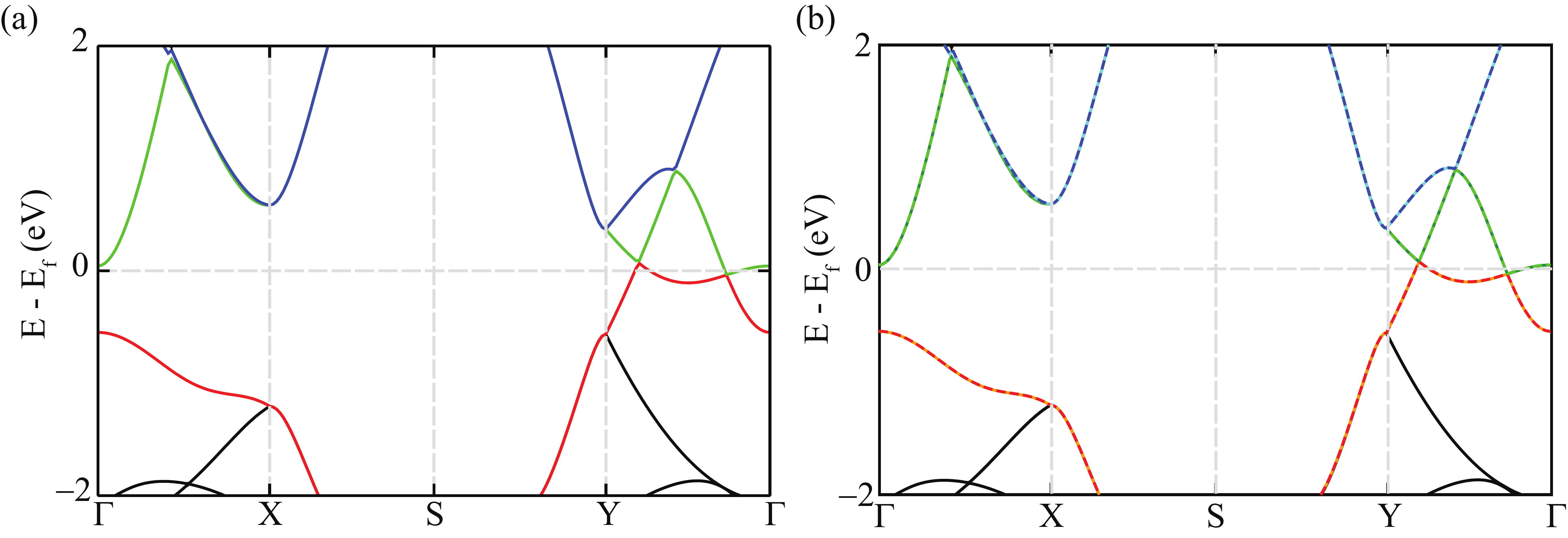} 
\caption{Electronic bandstructures of aluminane calculated under (a) PBE and (b) PBE with SOC functional.}
\label{figS2}
\end{figure}

\begin{figure}[!htbp]
\centering
\includegraphics[width=1.0\columnwidth]{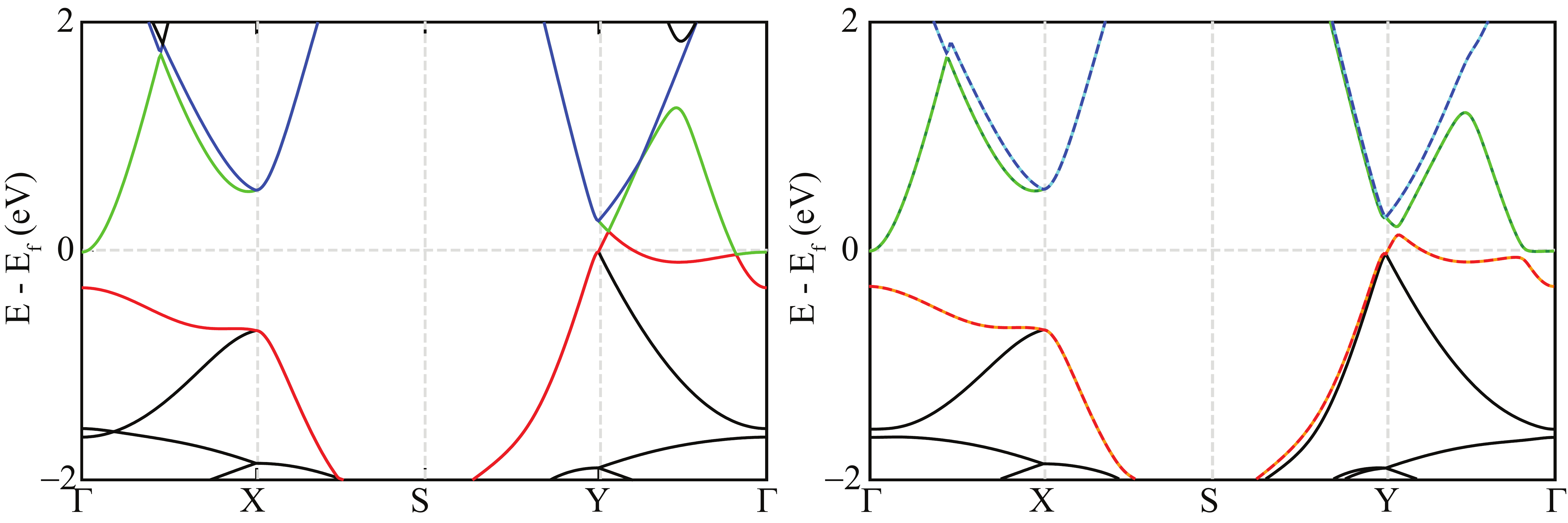} 
\caption{Electronic bandstructures of indinane calculated under (a) PBE and (b) PBE with SOC functional.}
\label{figS3}
\end{figure}

\begin{figure}[!htbp]
\centering
\includegraphics[width=1.0\columnwidth]{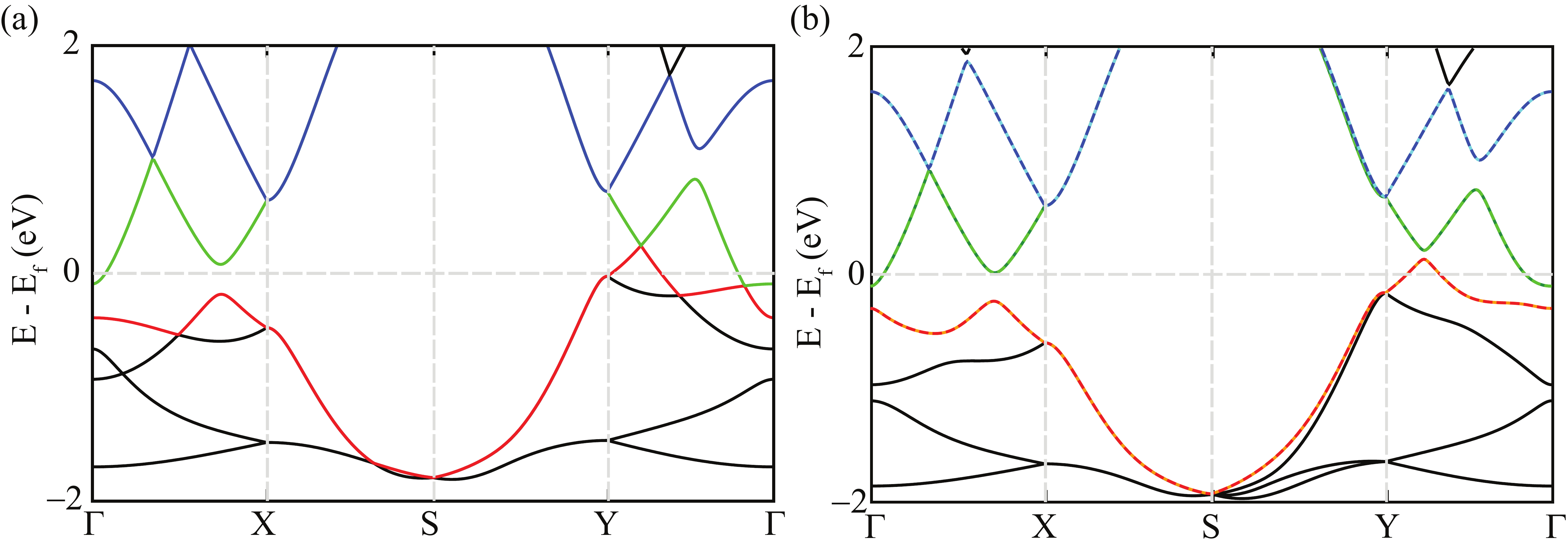} 
\caption{Electronic bandstructures of thalinane calculated under (a) PBE and (b) PBE with SOC functional.}
\label{figS4}
\end{figure}